\documentclass[12pt]{article}
\usepackage{epsfig}
\usepackage{graphics}
\usepackage{ulem}

\def\be{\begin{equation}}
\def\ee{\end{equation}}

\def\vareps{\varepsilon}
\def\pmcond{\stackrel{\scriptstyle{+}}{\scriptstyle{(-)}}}
\def\mpcond{\stackrel{\scriptstyle{-}}{\scriptstyle{(+)}}}
\begin{document}
\begin{center}
\date{today}
\Large{\textbf{Density of states of helium droplets}}\\
\vspace{0.5cm}
\large{Klavs Hansen}\footnote{email: klavs@physics.gu.se\\
phone:+46 (0)31 772 3432\\ FAX: +46 (0)31 772 3496}\\
{\it Department of Physics, G\"{o}teborg University, SE-412 96 G\"{o}teborg, Sweden}\\
\vspace{0.5cm}
\large{Michael D. Johnson, Vitaly V. Kresin}\\
{\it Department of Physics and Astronomy, \\University of Southern
California, Los Angeles, California 90089-0484, USA}\\
\vspace{1cm} \today
\end{center}
\vspace{1cm}

\begin{abstract}
Accurate analytical expressions for the state densities of liquid
$^4$He droplets are derived, incorporating the ripplon and phonon
degrees of freedom. The microcanonical temperature and the ripplon
angular momentum level density are also evaluated. The approach is
based on inversions and systematic expansions of canonical
thermodynamic properties.
\end{abstract}

\section{Introduction}

Important dynamical processes in finite systems such as nuclei,
polyatomic molecules, nanoclusters, atomic clouds, droplets
frequently turn out to be statistical in nature:
evaporation/fragmentation, radiation, emission of electrons,
equilibration between internal degrees of freedom or between host
and solvent molecules. When such a system is thermally isolated,
e.g. when flying in a beam or suspended in a trap, the proper
statistical-mechanics treatment is that of the microcanonical
ensemble where the energy $E$ is fixed and not the temperature of an external
heat bath. The density of
states, or level density, $\rho(E)dE$ represents the number of
quantum states between energy $E$ and $E+dE$. For separable degrees of freedom
this is number of normal mode combination
such that their energies add up to a total internal energy lying in this interval.
The function plays a crucial role in the thermal
description of microcanonical systems. For low excitation
energies $\rho(E)$ can be represented
by a sum of delta functions, corresponding to excitations of a only a few of the
individual modes, but for even moderate excitation
energies the density of these delta functions becomes so large that
is is well described as a continuous function of energy.
In this situation it is most convenient to use a
density smoothed over the discreteness of the energy levels. In addition to energy systems described
with the microcanonical ensemble have a conserved total angular momentum, so the
correspondingly resolved density of states, $\rho(E,J)$, is often of
relevance.

Free liquid helium nanodroplets \cite{gen1,gen2} represent an
interesting system for a statistical treatment. One reason is
that helium is the only element which cannot be described in terms
of classical dynamics for any internal degrees of freedom under the
experimental conditions used to study the droplets. This
makes the system interesting in its own right. For example, "magic
number" maxima in the size distributions of small $^4$He clusters
have been shown to correlate with the ability of the cluster to
accommodate elementary excitation modes \cite{Toennies-magic}. A
second reason is the use of the droplets as micro-cryostats used to
investigate other clusters and molecules. Evaporative cooling
generates internal energies corresponding to temperatures of
$\approx$ 0.4 K and is used to thermalize impurities to this
otherwise unreachable temperature for gas phase molecule and cluster
beams. Understanding these processes requires accurate
density-of-states expressions for the elementary excitation
spectrum.

The calculation of level densities requires that the excitation
spectrum is known. At low temperatures, the relevant normal modes of
$^4$He$_N$ clusters within the liquid drop model are ripplons which
are quantized capillary surface waves, and phonons which are
quantized bulk compression waves. For large droplets these modes are
separable to a good approximation \cite{Tamura-1996}, a fact that
greatly facilitates a statistical analysis of the excitation
spectrum. For a spherical droplet, both ripplon and phonon modes
possess well-defined eigenvalue spectra characterized by angular
momentum for ripplons, and angular momentum and mode index for
phonons.  A calculation of the total density of states requires
enumeration of all possible normal mode combinations, with
individual energies and angular momenta adding up to a given total
$E$ and $J$.

A leading-order evaluation of $\rho(E)$ for ripplons was carried out by
Brink and Stringari \cite{Brink+Stringari-1990}. Subsequently,
Lehmann \cite{Lehmann-2003-1}
presented a comprehensive discussion of the densities of states for ripplons
and phonons computed by direct numerical counting, and showed that the
resultant plots of the logarithm of the level densities
could be well parameterized by polynomial fits. These fits were
then used to calculate other thermodynamic functions and to
analyze droplet cooling with angular momentum conservation
constraints \cite{Lehmann-2003-2,Lehmann+Doktor-2004}.

In this paper, we show that accurate density of
states functions can be obtained by analytic evaluation. This is
appealing in its own right, as the calculations take advantage of
several elegant and generally useful tools from the literature.
In addition, having analytic expressions for various types of
elementary excitations provides a systematic method for treating
situations where several types of normal modes are excited
simultaneously, or when the spectrum of elementary excitations is modified.

The plan of the paper is as follows.  In Section \ref{ripplons} we
calculate the ripplon density of states as a function of energy,
Section \ref{ripplons-L} considers its angular momentum dependence,
Section \ref{phonons} is devoted to phonon excitations, Section \ref{phonons-L}
to the angular momentum of phonons, and Section
\ref{cumulative} to the total $\rho(E)$ function. Section
\ref{conclusions} comments on the similarity between the spectra
considered here and those of multielectron bubbles in bulk liquid
helium, and presents a summary.

\section{Ripplon density of states} \label{ripplons}

As mentioned above, ripplons are quantized waves on the droplet surface.
For a spherical liquid drop, the elementary excitation spectrum is given by
\cite{L&L-1}
\be\label{ripplon1}
\varepsilon _\ell   = \hbar \omega _0 \sqrt {\ell (\ell  - 1)(\ell  + 2)}.
\ee
Here  $\ell  \ge 2$ is the angular momentum quantum number of
the wave and
\be\label{ripplon2}
\omega _0  = \sqrt {\frac{\sigma _t }{DR3 }}  = \sqrt {\frac{4\pi \sigma _t }{3m_a N}},
\ee
where $\sigma_t$ is the coefficient of surface tension, $D$ the mass
density, $R$ the droplet radius, $m_a$ the atomic mass, and $N$ the
number of atoms in the droplet. If the parameters of bulk liquid
helium are used, we have $\hbar \omega _0  \approx \left( {3.8/\sqrt N } \right)$K
in temperature units \cite{Lehmann-2003-1}.  Below, the ripplon energy will be
expressed
in dimensionless units scaled to the quantity $\hbar \omega _0$.
Each mode has a degeneracy of $(2\ell  + 1)$.

\subsection*{Canonical approximation} \label{ripplons-can}

A first approximation to the level density can be derived in the
canonical ensemble picture, where it is assumed that the system
possesses a definite temperature $T$, and the system's internal
energy is associated with its most probable value. The energy
density of states of a finite system is then given by
\cite{Brink+Stringari-1990,Bethe1937,Ericson1960,Andersen2001}

\be\label{can}
\rho(E) = \frac{e^S}{\sqrt { - 2\pi (\partial
E/\partial \beta )} },
\ee
where $\beta  \equiv \left( {k_B T}\right)^{-1}$ and
\be\label{entropy}
 S = \beta E + \ln Z
\ee is the entropy; $Z$ is the canonical partition function. In the following we will
use units where $k_B = 1$. The square root appearing in the equation
involves the heat capacity and appears because the canonical entropy
includes an approximately gaussian integral over the thermally
populated states with a width given by the heat capacity and the
temperature, see, e.g., Ref. \cite{Andersen2001}.

Since the ripplon elementary excitations are bosons we have
\be\label{ripplonZ}
\ln Z =  - \sum \limits_{\ell =2}^{\ell_{max}} (2\ell+1)
\ln\left( 1-e^{-\beta\vareps_{\ell}} \right).
\ee
The canonical thermal energy of the ripplon ensemble is
\be\label{thermalE}
E =  - \partial (\ln Z) / \partial \beta.
\ee

To leading order, we can replace the sum in Eq. (\ref{ripplonZ}) by an
integral from zero to infinity, and approximate the energy
eigenvalues (\ref{ripplon1}) by $\varepsilon _\ell   \approx \ell ^{3/2} $.
The integral then straightforwardly evaluates to
\be\label{ripplonZapp}
\ln Z = \Gamma \left( \frac{7}{3} \right)
\zeta \left(\frac{7}{3} \right)
\beta ^{ - 4/3}  = 1.685\beta ^{ - 4/3},
\ee
and from Eq. (\ref{thermalE}) the (dimensionless) energy is
\be\label{thermalE2}
E = 2.247\beta ^{ - 7/3}.
\ee

Assembling everything into Eq.(\ref{can}) and expressing the answer in
terms of the energy, we find
\be\label{rho-ripplon1}
\rho _{rip} (E) \approx 0.311E^{ - 5/7} \exp (2.476E^{4/7} ),
\ee
which is the same answer as in Ref. \cite{Brink+Stringari-1990}.

\subsection*{Microcanonical ensemble} \label{ripplons-micro}

The above calculation can be improved in two places. One obvious refinement is to
evaluate the sum (\ref{ripplonZ}) with
greater care and to use more precise eigenvalues. A deeper
conceptual question is how to compute thermodynamic quantities for
a finite isolated system for which the total internal energy is a
conserved quantity and not an expectation value and the use of a
"temperature" must be carefully defined. A thorough discussion
was given by Andersen et al. in Ref. \cite{Andersen2001} with the conclusion that
the convenient canonical formalism may be retained, but the
canonical expression for the energy (\ref{thermalE}) must be corrected as
follows:
\be\label{miccanT}
E =  - \partial \left(\ln Z \right)/\partial \beta - \beta ^{ - 1}.
\ee
Here $E$ is the fixed excitation energy of the system and $\beta$ is
understood as the "microcanonical temperature" defined as
\be\label{defineT}
\beta \equiv \partial [\ln \rho (E)]/\partial E.
\ee

The procedure taken is as follows. First, the sum in Eq.
(\ref{ripplonZ}) is calculated using the first three terms of the
Euler-Maclaurin summation formula \cite{HandbookMath}.  With the form of the spectrum
given in Eq.(\ref{ripplon1}), this formula becomes
\begin{eqnarray}\label{F-Euler-Mac}
- \ln Z
&=& \int\limits_2^\infty  {(2\ell  + 1)\ln \left( {1 - e^{ - \beta \varepsilon _\ell
 } }
\right)d\ell }
+ \frac{5}{2}\ln \left( {1 - e^{ - \beta \varepsilon _2 } } \right)\\
\nonumber
&-&
\frac{1}{12} \left.  \frac{d}{d\ell} \left[ (2\ell  + 1)
\ln \left( 1 - e^{ - \beta \varepsilon _{\ell}}  \right) \right]
\right|_{\ell  = 2}+...
\end{eqnarray}
The upper limit, $\ell_{max}$, has been set to infinity as before. The
actual value is on the order of $\ell_{max} \approx {2\pi
R}/\lambda_{min} \approx {2\pi R}/(2d)$, where $\lambda$ is the
wavelength and $d$ is the interatomic distance \cite{Tamura-1996}.
In the liquid drop approximation ($R=N^{1/3} d/2 $) one then has
$\ell_{max} \approx \pi N^{1/3}/2$. In view of Eqs. (\ref{ripplon1},
\ref{ripplon2}) this yields a size-independent ripplon Debye
temperature of $\vareps_{max} \approx 7.5$ K. Using this value to
estimate the error in $\ln Z$, the leftover terms are found to be on
the order of $\left(\beta \vareps_{max} /4 -7\ell_{max} /6\right)
\exp(-\beta \vareps_{max}).$  For $T=1K$ this is a relative
contribution to $\ln(Z)$ of less than $10^{-2}/N^{1/3}$ which will
be ignored.

\begin{figure}
\begin{center}
    \includegraphics [angle=270,width=1.0\textwidth]{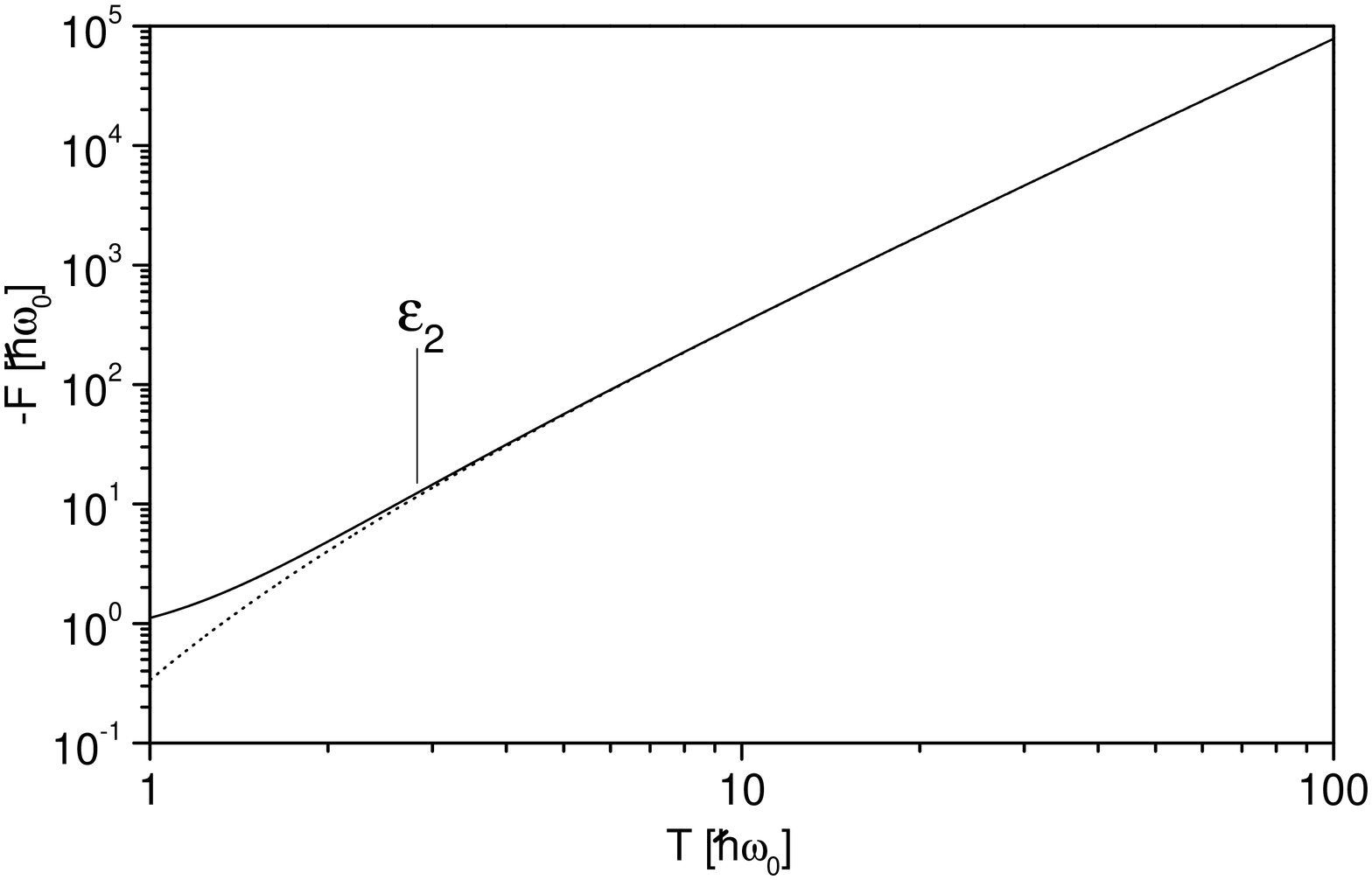}
   \caption{The (negative of) the ripplon free energies, calculated with
Eq.(\ref{Z(beta)}) (dotted line) and the
   summation in Eq.(\ref{ripplonZ}) (full line) which is exact apart from setting the upper
limit to infinity.
   The temperature corresponding to the energy of the lowest excitation,
$\vareps_2$, is indicated.}
    \label{F-comp}
  \end{center}
\end{figure}

A tedious calculation of Eq.(\ref{F-Euler-Mac}) involving expansions
of exponentials in powers of $\beta$ results in \cite{tobepublished}
\be\label{Z(beta)}
\ln Z = 1.685\beta ^{ - 4/3}  +
{\rm{0}}{\rm{.639}}\beta ^{ - 2/3} - \frac{{349}}{{96}} +
\frac{7}{3}\ln (2\sqrt 2 \beta ) + ...
\ee
The first term coincides
with Eq. (\ref{ripplonZapp}), and the rest are finite-size and
spectral corrections.  Note that all the numerical coefficients
derive from explicit expressions involving special functions.  The
expansion Eq.(\ref{Z(beta)}) has been checked against a numerical
sum. The comparison is shown in Fig. \ref{F-comp} for Helmholtz' free energy,
$F = -T \ln(Z)$. Already at
temperatures where $T$ is equal to the lowest excitation energy
$\varepsilon_2=\hbar \omega_0 \sqrt{6}$, the free energy is well
represented by the above expression. At higher energies the
agreement improves monotonically.

Knowing the partition function, we can now use Eq. (\ref{miccanT})
to determine the relation between the microcanonical energy and
temperature:
\be\label{E(beta)}
E = 2.247\beta ^{ - 7/3}  + {\rm{0}}{\rm{.426}}\beta ^{ - 5/3}  -
\frac{{10}}{3}\beta ^{ - 1}.
\ee
Again, the first term reproduces Eq. (\ref{thermalE2}).

In order to proceed with the calculation of the entropy, the heat
capacity and the level density in Eq.(\ref{can}), we need to invert
the relation (\ref{E(beta)}) which expresses $E(\beta)$ to get
$\beta (E)$. This is done by the iterative method of successive
approximations.  The result is an expansion for $\beta^{-1}$ in
powers of $E^{-2/7}$,
\be\label{caloric1}
\beta^{-1} \equiv 0.7069 E^{3/7} -0.07239
E^{1/7} + 0.7212 E^{-1/7}+...,
\ee
where the coefficients are
calculated from those entering Eq.(\ref{E(beta)}). Now the prefactor
and the exponent in Eq.(\ref{can}) can be evaluated, using Eqs.
(\ref{entropy}),(\ref{Z(beta)}),(\ref{E(beta)}), and $\beta(E)$,
finally yielding

\be\label{ripplonLD}
\rho _{rip} (E) = 0.205E^{ - 12/7} \exp \left(
{2.476E^{4/7} + 0.507E^{2/7} } \right).
\ee
Let us emphasize
again that all the numerical coefficients encode analytical
expressions.

\begin{figure}
\begin{center}
\includegraphics [angle=270,width=1.0\textwidth]{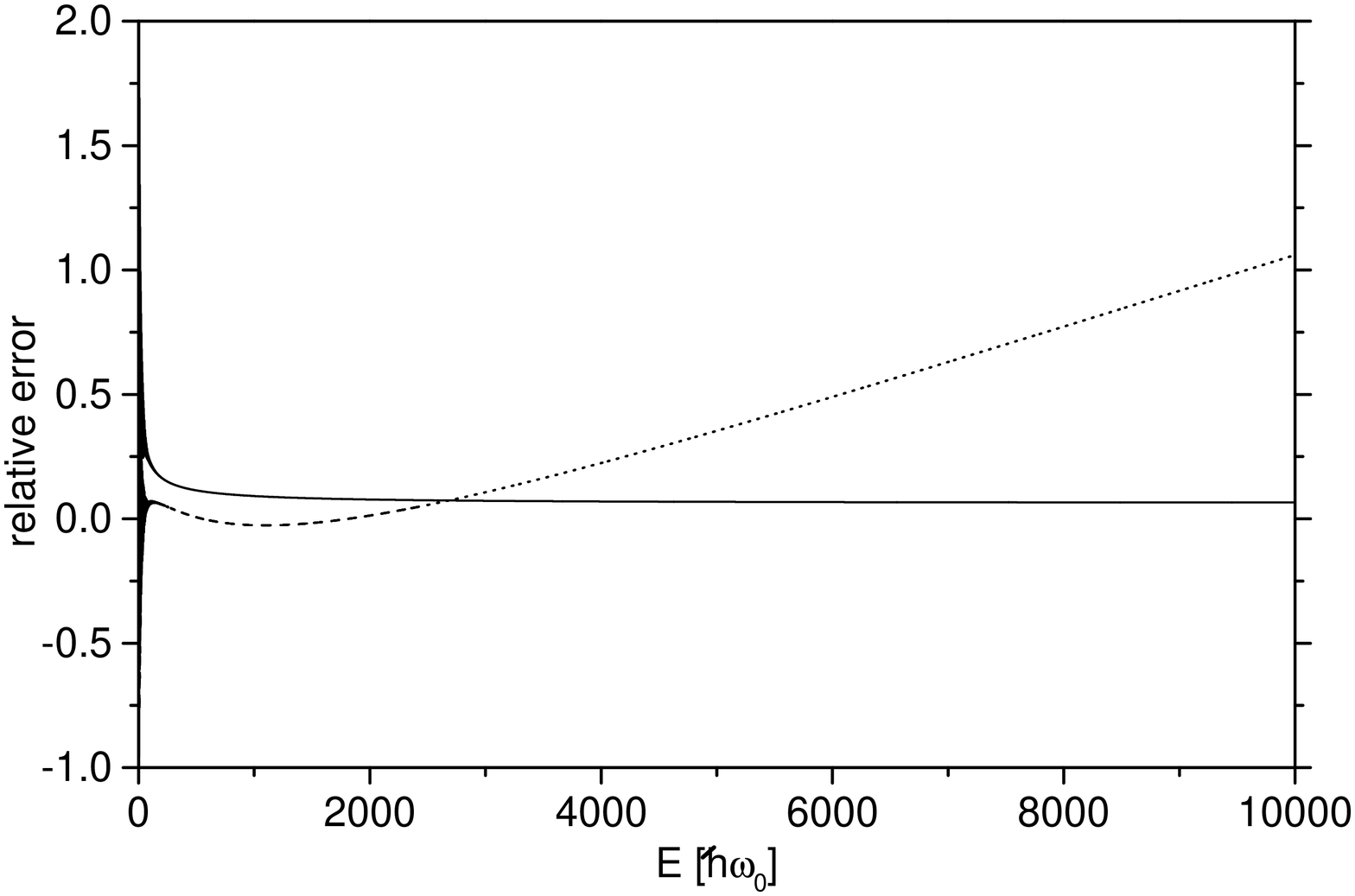}
   \caption{Comparison of the ripplon level densities calculated according to
   Eq.(\ref{ripplonLD}) (full line) and Ref.\cite{Lehmann-2003-1} (dashed for $E < 2500$, dotted line
   for $E> 2500$). The fit in Ref.\cite{Lehmann-2003-1} was limited to
   energies between 50 and 2500, in the reduced units used here and is
   calculated as the derivative of the numerical fit to the integrated level
   density. The expressions have been divided by the exact
   Beyer-Swinehart result, causing the oscillatory behavior at low energy,
   and the curves plotted are the logarithms of these ratios. The curve of
   Ref.\cite{Brink+Stringari-1990} (not shown) is around 3.}
\label{rho-comp}
  \end{center}
\end{figure}

Eq.(\ref{ripplonLD}), which is the main result of this section, may
be compared with the canonical approximation (\ref{rho-ripplon1}),
an exact numerical count carried out with the help of the
Beyer-Swinehart algorithm \cite{BS}, and the form  written down in
Ref. \cite{Lehmann-2003-1} as an empirical fit to the numerical
count in the interval E=50-2500.  Fig. \ref{rho-comp} shows such a
comparison, and demonstrates that the analytical expression gives an
excellent representation of the exact result \cite{footnote}.

\section{Ripplon angular momentum density}\label{ripplons-L}

The next step is to generalize the ripplon state density to a
function which is not only energy- but also angular
momentum-resolved. This problem has been comprehensively
studied in nuclear physics \cite{Bethe1937,Ericson1960}. One way of
visualizing the net angular momentum of a large distribution of
excitations with varying $(\ell ,\ell _z )$ is as the result of
random angular momentum coupling, in which case the central limit
theorem applies and one expects to find a normal distribution.
Indeed, the above references show that $\rho(E,J)$ is essentially a
product of $\rho(E)$ and a Gaussian factor:
\be\label{rho(E,J)}
\rho (E,J) = \rho (E)\frac{2J + 1}{2(2\pi )^{1/2} \sigma^3}
e^{ - \frac{J(J + 1)}{2\sigma^2 }}.
\ee
It is permissible here to replace $J(J+1)$ by $(J+\frac{1}{2})^2$.

It is still necessary to establish the variance $\sigma^2$. An
elegant way to do this to leading order by means of an extended
grand canonical distribution is described in Bethe's review
\cite{Bethe1937}, where the method is applied to a system of
non-interacting fermions in a spherical potential box.  Here we
follow the same procedure for a system of bosonic ripplon
excitations.

The idea is first to evaluate the projection $M$ of the
net angular momentum $\vec J$ of the droplet in terms of the
contributions of individual normal modes at temperature $T$.  The fact
that $M$ is a conserved quantity is accounted for by a separate
Lagrange multiplier, or "chemical potential" $\gamma$, such that

\be\label{gamma}
\gamma = - \partial S / \partial M
\ee
($S$ is the entropy). We can calculate $M$ directly by summing over
all modes:

\be\label{M}
M = \sum\limits_{\ell  = 2}^\infty  {\sum\limits_{m =  - \ell
}^\ell  {\frac{m}{{e^{\beta \varepsilon _\ell   - \gamma m}  -
1}}} }  \approx \int\limits_0^\infty  {d\ell } \int\limits_{ -
\ell }^\ell  {\frac{{m \cdot dm}}{{e^{\beta \ell ^{3/2}  - \gamma
m}  - 1}}}
\ee
(in reduced energy units).  Expanding the integrand to first order
in $\gamma$ \cite{Bethe1937}, we find

\be\label{M-2}
M = \frac{{20}}{{27}}\Gamma \left( {\frac{5}{3}} \right)\zeta
\left( {\frac{5}{3}} \right)\gamma \beta ^{ - 4/3},
\ee
and Eq. (\ref{thermalE2}) allows us to express the result in terms of the
droplet energy. To leading order we have:
\be\label{gamma2}
\gamma  = 1.776ME^{ - 8/7}.
\ee
Now we can use Eq. (\ref{gamma}) with Eq. (\ref{gamma2}) to obtain the entropy
variation:
\be\label{S(E,M)}
S(E,M) = S(E,0) - M^2 /(2\sigma ^2 )
\ee
with
\be\label{M-sigma} \left( {2\sigma ^2 } \right)^{ - 1}  = 0.888E^{ -8/7}.
\ee

The second term in Eq. (\ref{S(E,M)}) leads to a normal distribution
in $M$. The distribution in $J$ can be shown to have the same
variance \cite{Bethe1937,Ericson1960}.  Therefore Eqs.
(\ref{rho(E,J)}) and (\ref{M-sigma}) define $\rho_{rip}(E,J)$.

The numerical evaluation of the rotational density of states in Ref.
\cite{Lehmann-2003-1} led to essentially the same form of the state
density function, with the factor corresponding to
$(2\sigma^2)^{-1}$ fitted as $0.868E^{-8/7}+0.964E^{-13/7}$, which
affirms the analytical result (\ref{M-sigma}): the factors deviate
by less than 2\% for $E=$100-2500.

A shorter estimate of the variance is illustrative. The number of
quanta in one mode $(\ell,m)$ is on the order of $T/\vareps_{\ell}$
for levels up to $\vareps_{\ell}\simeq T$ and zero for higher
quantum energies. The total number of excited quanta is then

\be n \approx \sum_{\ell=2}^{T^{2/3}}(2\ell +1)T /\ell^{3/2} \approx
4 T^{4/3}, \ee
where the sum was approximated by an integral and $T$
is written in terms of the $\omega_0$ unit. With the
energy-temperature relation (\ref{E(beta)}) we get the leading order
value for energy per quantum $\langle e \rangle = E/n = 2.247T/4,$
and from this an average of $\langle \ell \rangle = \langle e
\rangle^{2/3} = T^{2/3} (2.247/4)^{2/3}.$ The standard deviation
$\sigma$ of $\ell$ is then, according to the 'random walk' argument
used above, $\sigma = \sqrt{n} \langle \ell \rangle$. Inserting the
calculated $\langle \ell \rangle$ and expressing the result in terms
of the total energy, one has

\be \label{rip-sigma} \left(2 \sigma^2 \right)^{-1} =
2^{-1/3}\left(2.247 \right)^{-4/21} E^{-8/7} =0.68 E^{-8/7}, \ee
in
surprisingly sensible agreement with the above result.

One may seek to describe the angular momentum distribution in the
language of a rotational energy and a moment of inertia $I$,
associating \cite{Ericson1960} the exponential in Eq. (\ref{rho(E,J)}) with a Boltzmann
factor involving $\beta \hbar ^2 J(J + 1)/\left( {2I} \right)$,
i.e., $I = \hbar ^2 \beta \sigma ^2 $ .  Using the
canonical-ensemble results, Eqs.(\ref{thermalE2}) and (\ref{M-sigma}), we can
express the
"ripplon moment of inertia" in terms of the ripplon excitation
energy (in reduced units):
\be\label{I}
I = 0.797E^{5/7}.
\ee

\section{Phonon density of states}\label{phonons}

Surface ripplons are the lowest-temperature droplet excitations;
bulk phonons appear next. These are compression sound waves which
arise as solutions of the wave equation within the volume of the
spherical drop.  As such, their energies are given by

\be\label{phonon-disp}
\vareps _{n,\ell }  = \hbar uk_{n,\ell }
\ee
where $u$ is the speed of sound and the wave number $k_{n,\ell}$
is determined by the boundary condition at the surface. If the Dirichlet
boundary condition is adopted \cite{Tamura-1996, Lehmann-2003-1}, then
$k_{n,\ell}= a_{n,\ell}/R$,
where $a_{n,\ell}$  is the {\it n}th root of the $j_\ell$ spherical Bessel function.
For a free surface, a more appropriate boundary condition is the Neumann one, in which
case $k_{n.\ell }= a'_{n,\ell}/R$, with $a'_{n,\ell}$ the root of the Bessel function
derivative, $j'_\ell$. The energy scale is set by the longest wave length, i.e.,
$k \sim \pi /R$, so we can express phonon energies in units of
\be\label{ph-cutoff}
\tilde \varepsilon  = \hbar u\pi /R,
\ee
which works out to $\tilde \varepsilon  = \left( {25.5N^{ - 1/3}}
\right)$K in temperature units if the speed of sound in bulk
$^{4}$He is used \cite{Lehmann-2003-1}. The leading-order behavior
of the phonon state density can be determined in a straightforward
way by invoking the standard expression for the Debye heat capacity
(per unit volume) of bulk phonons:
\be\label{Cvbulk}
C_{bulk}  = \frac{2\pi ^2 }{15}\frac{k_B^4 }{\hbar^3 u^3}T^3.
\ee

Multiplying this by $4\pi R^3 /3$ and using the fact that (in the
canonical framework) $C = \partial E/\partial T$ and $\left( {k_B T}
\right)^{ - 1}  = \partial S/\partial E$, we can use integrations to
express $S$ in terms of $E$.  Then, from Eq.(\ref{can}) we find that
to first order, $\ln \rho _{ph} (E) \approx S(E) = 3.41E^{3/4}$.
This matches the leading term of the fit to a direct numerical count
in Ref. \cite{Lehmann-2003-1} which is $3.331 E^{3/4}$.

The Debye temperature for phonons in liquid $^4$He is $\approx$ 25 K
\cite{Woods+Cowley-1973}, corresponding to a total phonon thermal
energy (from Eq.\ref{phononE}) of $\approx 1000 N$ K. We can therefore use
the low temperature approximation throughout.

The prospect of refining the calculation by analytically evaluating
a statistical sum over the precise spectrum (\ref{phonon-disp}) may
seem bleak, as the Bessel function roots which "contribute in an
essential manner… are just the zeros for which the usual formulae
(like McMahon's expansion) are bad approximations"
\cite{Baltes+Hilf-book}.  However, rescue comes from an elegant
mathematical framework known as the Weyl expansion
\cite{Baltes+Hilf-book,Baltes+Hilf-1972,Brack+Bhaduri}.  It
provides a systematic expression for the smoothed density of
eigenmodes in a finite cavity in terms of volume, surface, and
curvature terms.  As described in the above references, this is a
very general theory, valid for both scalar and vector wave
equations, and applicable to a wide variety of physical phenomena.

Ref. \cite{Baltes+Hilf-1973} applied this formalism to the specific
heat of metal nanoparticles.  The finite-size correction to the
specific heat (\ref{Cvbulk}) derived there is immediately usable for
our droplet problem:

\be\label{Cvdroplet}
C = C_{bulk} \pmcond \frac{9\zeta
(3)}{4\pi}\frac{k_B^3}{\hbar^2 u^2 } \frac{T^2}{R} +
\frac{1}{6}\frac{k_B^2}{\hbar u}\frac{T}{R^2}.
\ee
The + sign applies to the Neumann and the - sign to the Dirichlet boundary
conditions. Although we focus on the Neumann condition, the
Dirichlet case will be included for completeness.

We now follow almost the same sequence as in the bulk limit
described above:  Eq. (\ref{Cvdroplet}) is multiplied by the droplet
volume and integrated once to obtain (with $E$ and $T$ in units of
$\tilde \varepsilon$)
\be\label{phononE}
E(T)= \frac{{2\pi ^6 }}{{45}}T^4  \pmcond \zeta
(3)\pi ^2 T^3  + \frac{{\pi ^2 }}{9}T^2,
\ee
and a second time to
obtain $S(T)$ as $S = \int_0^T (C/T')dT'$. The first function is
inverted by iteration to yield
\be
T(E) = 0.391 E^{1/4} \mpcond 0.069 -0.006E^{-1/4}.
\ee
[Calculating $T(E)$ instead of $\beta(E)$ is more convenient in this
case.] Eq. (\ref{can}) is then used to obtain the density of states.
The calculation is done to the first three orders in $E$, in
correspondence to the three terms in the heat capacity expansion
(\ref{Cvdroplet}). The microcanonical correction (\ref{miccanT}) in
the present case turns out to contribute only in the next order of
smallness. The result of the calculation is as follows:
\be\label{rho-ph}
\rho _{ph} (E) = AE^{ - 5/8} \exp \left(
{3.409E^{3/4} \pmcond 0.908E^{1/2}  + 0.482E^{1/4} } \right)
\ee
Once again, the + sign is for the Neumann boundary condition on the
phonon wave at the droplet surface and (-) for the Dirichlet
condition. Using the bulk canonical heat capacity in
Eq.(\ref{Cvdroplet}) gives a pre-exponential factor of $A=0.26$.

\begin{figure}
\begin{center}
\includegraphics [angle=270,width=1.0\textwidth]{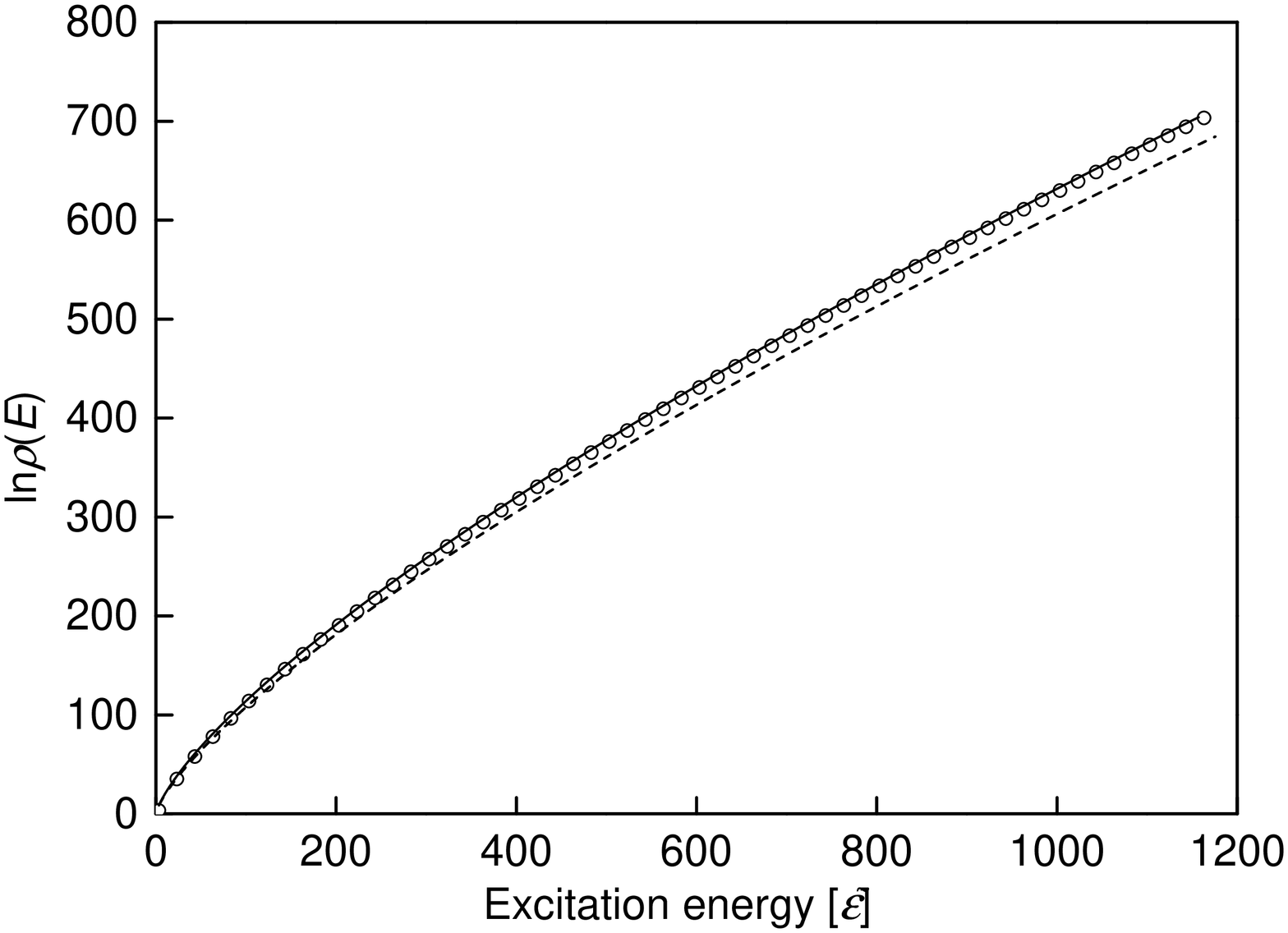}
\caption{Phonon level densities calculated according to the exact
Beyer-Swinehart count (open circles), Eq.(\ref{rho-ph}) (full line),
and the level density derived from the bulk Debye heat capacity,
i.e. corresponding to Eq.(\ref{rho-ph}) but including only the first
term in the exponential (dashed line).} \label{phonon-comp1}
\end{center}
\end{figure}

\begin{figure}
\begin{center}
\includegraphics [angle=270,width=1.0\textwidth]{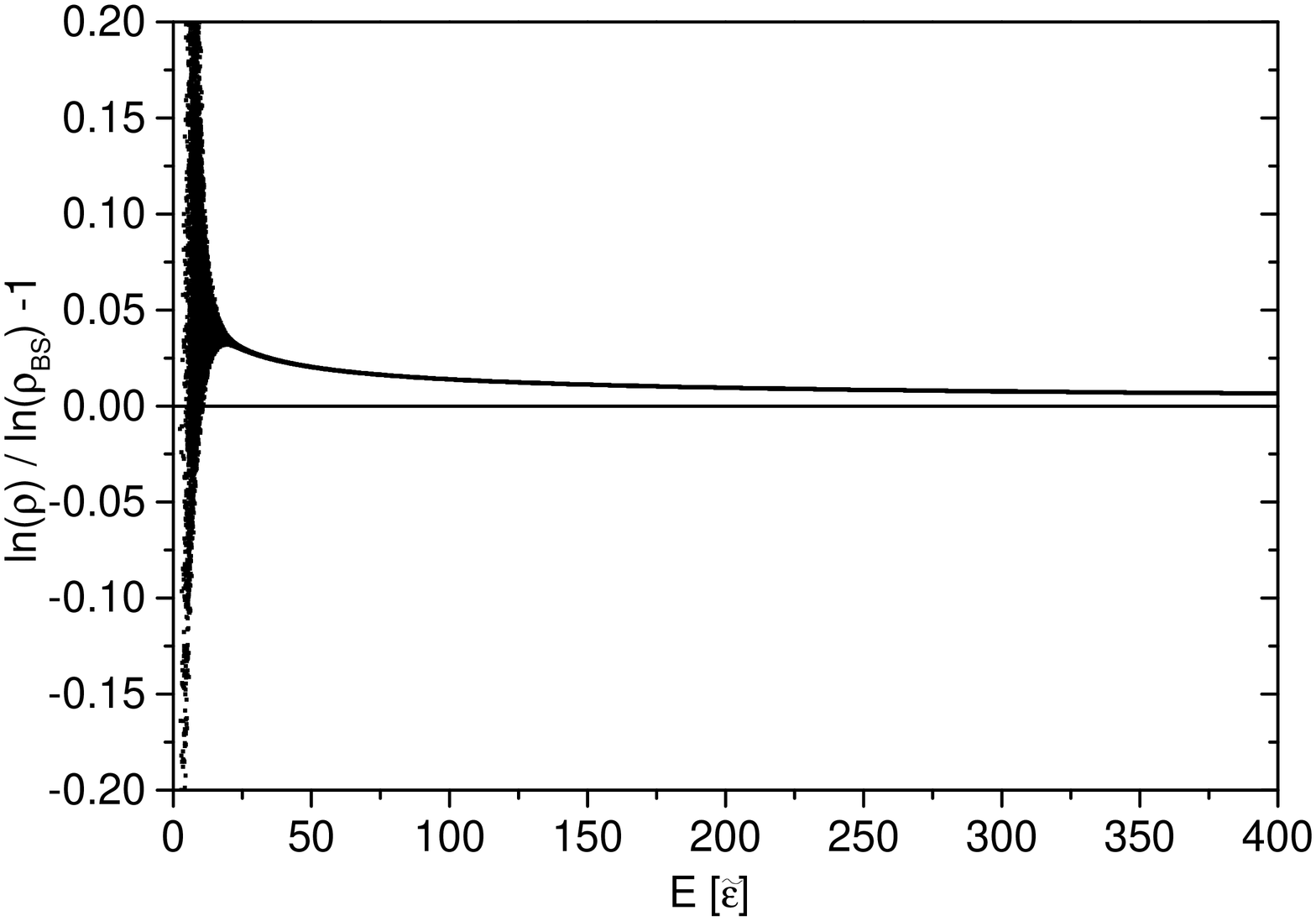}
\caption{A comparison of Eq.(\ref{rho-ph}) and the exact-count phonon level
density, showing essentially the relative
difference in the entropy of the phonon system in the two calculations.}
\label{phonon-comp2}
\end{center}
\end{figure}

Fig.\ref{phonon-comp1} compares the exact Beyer-Swinehart count for
the phonon spectrum with the full Eq.(\ref{rho-ph}) and with the
level density based on the bulk Debye heat capacity, Eq.
(\ref{Cvbulk}), i.e., where only the first term in the exponent is
present. Fig.(\ref{phonon-comp2}) shows a more detailed comparison
of Eq.(\ref{rho-ph}) and the exact-count phonon level density. We
find good agreement between analytical expression and the exact
computation, although not as good as for the ripplon case.

The estimate of the prefactor $A$ in Eq. (\ref{rho-ph}) cannot be
expected to be correct because it does not include higher-order
expansion terms in the exponent that would yield corrections of the
same order. A comparison with the numerical count suggests a
correction in the form of a factor $\exp(-0.62E^{0.2})$. Although
this correction is larger than the error found for the ripplon level
density, it is nevertheless still relatively small. An effective
value of $A \approx 0.05$ can be used for energies below 400.


\section{Phonon angular momentum density}\label{phonons-L}

A computation of the angular momentum resolved phonon level density
suffers from the difficulties with expressing the lowest Bessel
function eigenvalues with a simple functional form. In contrast to
$\rho_{ph}(E)$ there is, to our knowledge, no
solution in the literature for this problem. As will be clear from
the results presented in section \ref{cumulative} below, the
contribution to the level density from the phonons is minor compared
to that of the ripplons, and the required precision in the
calculation of the angular specified phonon contribution is
therefore correspondingly smaller. In this section we will make an
order of magnitude estimate, based on the leading order term of
McMahon's expansion of the roots of the Bessel functions
\cite{HandbookMath}. For the Neumann boundary condition the roots
are $(n+\ell/2-3/4)\pi \approx (n+\ell/2)\pi$. With the phonon energy scale
used, Eq.(\ref{ph-cutoff}), the quantum energies are thus
$n+\ell/2$. When states with energies up to $T$ are averaged, the
linear dependence of the quantum energy on $\ell$ gives an average
value of $\langle \ell \rangle \sim T$. Since also the
$n$-dependence is linear the constant of proportionality is on the
order of unity. The total number of states below energy $T$ is on
the order of $T^3$. Combining these estimates give, using the same
type of 'random walk' estimate as Eq. (\ref{rip-sigma}) for the
ripplons, that
\be
{(2\sigma_{ph}^2)}^{-1} \sim \frac{1}{T^3 \cdot T^2} =
\left(\frac{2\pi^6}{45} \right)^{5/4} E^{-5/4} \approx 100E^{-5/4}.
\ee
The ratio of the $\sigma$'s for the phonons and ripplons (here
denoted $\sigma_{rip}$) with the leading order terms in the caloric
curves Eqs.(\ref{E(beta)},\ref{phononE}) and the proper energy
scaling is

\be \frac{\sigma_{ph}}{\sigma_{rip}} \sim 0.002 \left( T[K]
\right)^{7/6} N^{1/6}. \ee

This is small compared to unity up to extremely large droplet sizes.
The conclusion that the width of the phonon angular momentum
distribution can be ignored holds very well, even if the estimate of
the width should be incorrect by as much as an order of magnitude.

\section{Combined level density}
\label{cumulative}

A helium droplet may have both ripplon and phonon oscillations
excited at the same time (and, at higher temperatures, rotons as
well \cite{gen1}). The coupling between these normal modes is weak
at bulk liquid surfaces \cite{Reynolds-1992} and in large droplets
\cite{Tamura-1996}, thus their energy contents may remain
independently defined for some length of time, and the individual
state densities will then come from Eqs. (\ref{ripplonLD}) and
(\ref{rho-ph}). The question of equilibration dynamics of
excitations in superfluid droplets and the relevant time scales is a
very interesting one, and has not yet been addressed in detail.
Below, we discuss an estimate of state densities in circumstances
when the ripplon and phonon excitations do achieve statistical
equilibrium.

In principle, the level density of combined excitations can be
calculated by direct summation, as described in Section
\ref{ripplons-micro}. This would be a very involved procedure,
because the ripplon and phonon quantum energies have different
dispersion relations and scale differently with size. Alternatively,
one can calculate the level density as a convolution. Also in this task
does one benefit from formulating the general problem in terms of
the microcanonical temperature. The convolution to be performed is

\be\label{convolute} \rho(E) = \int_0^E \rho_{rip}(E-\vareps)
\rho_{ph}(\vareps)d\vareps. \ee For not extremely large droplets the
largest part of the excitation energy resides in the ripplons.
Indeed, the ratio between the energies of the ripplon and phonon
subsystems is, canonically:

\be\label{energy-ratio}
\frac{E_{ph}}{E_{rip}} \approx 6.8\times
10^{-3} N^{1/3} \left(T[K] \right)^{5/3}.
\ee
(Temperature expressed in Kelvins.) It is
clear that for temperatures under 1 K (i.e., those which lie safely
below the Debye cut-off values specified above and below the onset
of roton modes) and droplets of up to several tens of thousands of
atoms in size, the phonon energy contents is a fraction of the
ripplon energy. Under these conditions one can treat the ripplon
degrees of freedom as a heat bath and calculate the phonon
contribution with an expansion of the integrand of Eq.
(\ref{convolute}) around some small phonon energy. We will use the
simplest choice of zero phonon energy, and to increase the precision
we expand the {\it logarithm} of the level density. Thus

\be
\rho(E) = \int_0^E \rho_{rip}(E) \exp\left[-\vareps \frac{d\ln[\rho_{rip}(E)]}{dE}
+\frac{1}{2} \vareps^2\frac{d^2\ln[\rho_{rip}(E)]}{dE^2}-...
\right]\rho_{ph}(\vareps) d\vareps.
\ee
The upper limit of the integral can be replaced by infinity without
serious loss of precision because the integrand peaks well below
this value. We recognize the first derivative in the exponential as
the microcanonical temperature $1/T$ of the ripplon system at energy
$E$, see Eq. (\ref{defineT}), and therefore have

 \be \rho(E) =
\rho_{rip}(E) \int_0^{\infty} e^{-\vareps /T} \rho_{ph}(\vareps)
\exp\left[ \frac{1}{2} \vareps^2
\frac{d^2\ln[\rho_{rip}(E)]}{dE^2}+...\right] d\vareps. \ee
The
second exponential in the integrand can be expanded, with the
integral of the first term yielding the phonon canonical partition
function at $T$, $Z_{ph}(T)$:
\be\label{conv1}
\rho(E) = \rho_{rip}(E) \left\{Z_{ph}(T)
+\int_0^{\infty} e^{-\vareps /T} \rho_{ph}(\vareps)
\left[\frac{1}{2} \vareps^2 \frac{d^2\ln[\rho_{rip}(E)]}{dE^2}
+...\right] d\vareps \right\}. \ee
To leading order and ignoring the
difference between the canonical and microcanonical temperatures,
this simplifies to
\be\label{conv2} \rho(E) =
\rho_{rip}(E)Z_{ph}(T)\left\{1 - \frac{C_{ph}}{2C_{rip}} - \frac{E_{ph}^2}{2C_{rip}T^2}+...
\right\}.
\ee
Hence the ratio of the term which is second order in $\vareps$ to the zero
order term in Eq.(\ref{conv1}) is approximately
\be
\frac{C_{ph}}{2C_{rip}}+ \frac{E_{ph}^2}{2C_{rip}T^2} = 6 \times 10^{-3}
(T[K])^{5/3}N^{1/3} + 4\times 10^{-6}(T[K])^{14/3}N^{4/3}.
\ee
For not excessively large or warm droplets
we can leave out the correction terms and thus have
\be
\label{conv3}
\rho(E) = \rho_{rip}(E) Z_{ph}(T),
\ee
where
$Z_{ph}(T)$ as stated above is the phonon canonical partition
function at the microcanonical ripplon temperature corresponding to
the ripplon energy $E$.

The exponential part of the phonon canonical partition function
can be calculated, e.g., by integration of the standard relation
Eq.(\ref{thermalE}) with the caloric curve in Eq.(\ref{phononE}).
This procedure does not determine the integration constant which
translates into a multiplicative constant on the total level
density, Eq. (\ref {conv3}). This constant, $c$, is approximately
the product of the pre-exponential from Eq.(\ref{rho-ph}) and the
prefactor that appears in Eq.(\ref{can}) (i.e., the value given by a
saddlepoint expansion of the phonon level density in the calculation
of the canonical particion function). The result is \be c \approx
\sqrt{2\pi T^2 C_{ph}}AE^{-5/8}, \ee where $C_{ph}$ is again the
phonon heat capacity. The leading order expressions for the phonon
parameters $C_{ph}(T), E_{ph}(T)$ give, taking into account the
different scaling of energies for phonons and ripplons, the total
level density \be\label{conv} \rho(E) =\rho_{rip}(E)\cdot
0.526N^{1/6} \exp\left( 0.04713 N^{-1/2} T^3 + 0.01317 N^{1/3} T^2
+0.1634 N^{1/6} T \right) \ee with the equation given in ripplon
energy units and $T=T(E)$ given by Eq.(\ref{caloric1}). The constant
of $0.05$ for the phonon level density pre-exponential, mentioned at
the end of Sec. \ref{phonons}, was used here also.

\begin{figure}
\begin{center}
\includegraphics [angle=270,width=1.0\textwidth]{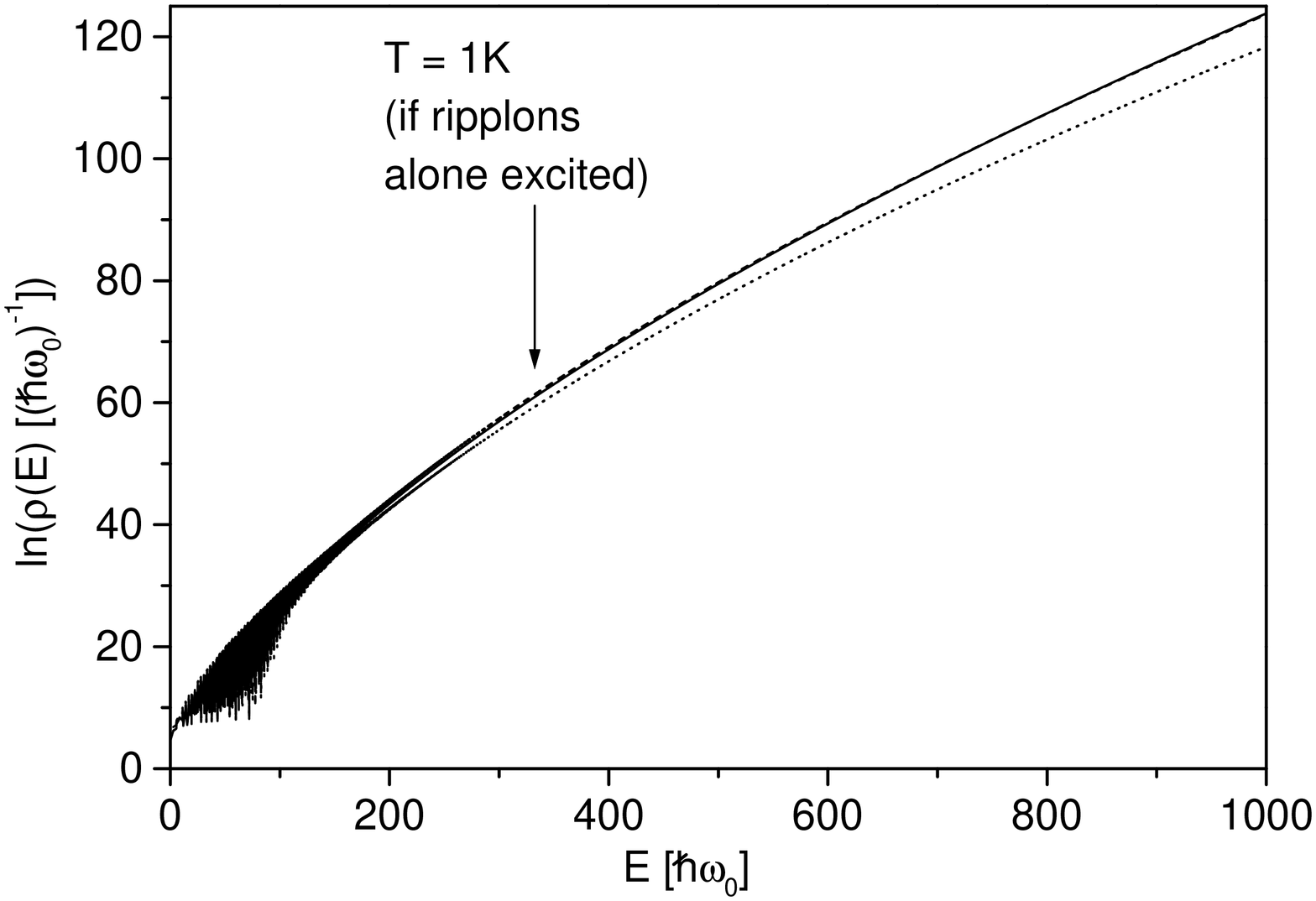}
\caption{The convoluted level densities for phonons and ripplons for droplet size
$10^3$. The numerical convolution is the full line, and the approximate result in Eq.(\ref{conv})
the, hardly discernible, dashed line.
The level densities for ripplons alone (dotted line) is given for reference. The arrow indicates the energy
content of the ripplon excitations at a temperature of 1 K.}
\label{convolution}
\end{center}
\end{figure}

This result is compared with numerical convolutions for $N=10^3$ in Fig.\ref{convolution}.
The numerical convolution was also calculated for $N=10^4$ with a similar result.

One remark about Eqs.(\ref{conv3},\ref{conv}) is in place: These
equations should only be used for calculations of microcanonical
properties. For the calculation of canonical properties one should
use the product partition function, $Z_{ripp,ph} = Z_{rip}Z_{ph}$. A
naive application of Eq.(\ref{conv}) in a calculation of the
partition function of the combined ripplon-phonon system will give a
divergent result at all temperatures. The origin of this divergence
is the breakdown at high excitation energies of the approximations
leading to the equation.

\section{Conclusions}\label{conclusions}
We have presented an analytical evaluation of the
statistical density of states functions of the elementary
excitations (surface ripplons and volume phonons) of isolated
liquid-drop helium nanoclusters.  These functions are expressed in
terms of microcanonically conserved quantities: energy and angular
momentum.  The obtained formulas accurately match numerically
computed curves as the energy level densities vary over $\sim 150-300$
orders of magnitude.

Other interesting helium systems to which the
results may be applicable include micron-sized superfluid fog \cite{Kim2002}
and multielectron bubbles in liquid helium.  The latter are
spherical voids inside bulk He, with a thin shell of electrons
lining the inner wall (see, e.g., Refs. \cite{Salomaa+Williams-1981, Tempere2001}
and references therein).  They can undergo small-amplitude shape oscillations,
i.e., ripplons, whose frequency under zero external applied
pressure has the form $\omega _\ell ^2  \propto (\ell ^2  -
1)(\ell  - 2)$ , which for large $\ell $
approaches the same form as the droplet ripplon dispersion, Eq. (\ref{ripplon1}).
This implies that the statistical mechanics of these bubbles should be
similar to that of nanodroplets.  One distinction is that the bubble are
submerged into a bulk helium thermal bath, therefore for them the canonical
ensemble treatment is rigorously correct and not just a convenient approximation.

Finally, it should be pointed out that the results obtained in the
present paper have a universal form and are expressed in terms of
dimensionless scaled energies, therefore they are generally applicable to
the statistics of droplets of various substances besides helium.

\section{Acknowledgments}
This work was supported by the Swedish National Research Council (VR),
the U.S. National Science Foundation
under Grant No. PHY-0245102, and a Lick fellowship to M.J.

\newpage

\newpage

\newpage

\newpage

\newpage


\begin{thebibliography}{99}
\bibitem{footnote} To resolve the numerical comparison, we empoyed two more digits
on the leading order term in the exponential
than written out in the text.
\end{thebibliography}
\end{document}